# Structure-guided taxonomic placement of divergent RNA viruses with ViraClass

Sheng Xu[1,2,*], Wenxuan Huang[1,2,*], Shutong Yue[2,3,*], Weiqiang Bai[1,2], Shiyang Feng[2], Xiaohan He[2], Bo Zhang[2], Qiantai Feng[1], Edward C. Holmes[4], Weifeng Shi[3,5,6,#], Siqi Sun[1,2,#]

[1]Research Institute of Intelligent Complex Systems, Fudan University, Shanghai, China
[2]Shanghai Artificial Intelligence Laboratory, Shanghai, China
[3]Ruijin Hospital, Shanghai Jiao Tong University School of Medicine, Shanghai, China
[4]School of Medical Sciences, The University of Sydney, Sydney, Australia
[5]Shanghai Institute of Virology, Shanghai Jiao Tong University School of Medicine, Shanghai, China
[6]School of Life Sciences and Biotechnology, Shanghai Jiao Tong University, Shanghai, China

[*]These authors contributed equally: Sheng Xu, Wenxuan Huang and Shutong Yue.
[#]Correspondence: shiwf@ioz.ac.cn; siqisun@fudan.edu.cn

## Abstract

Metatranscriptomic sequencing has expanded our knowledge of the RNA virosphere far more rapidly than novel viruses can be taxonomically classified. Taxonomic assignment above the family level is particularly difficult because the RNA-dependent RNA polymerase (RdRp) is often the only gene retained across RNA viruses yet exhibits little sequence similarity among highly divergent viruses. Here we show that RdRp protein structure retains taxonomic signal at evolutionary depths where RdRp primary sequence similarity has largely collapsed, and that the organization of this signal is consistent with the current ICTV hierarchy. Based on this, we developed ViraClass, a hierarchical framework for RNA virus taxonomic placement that uses RdRp structure for rank-by-rank assignment from phylum to genus, stopping at the deepest rank supported by confidence thresholds, and calibrated structural clustering for viruses that remain outside existing reference space. Across random-split, prospective and taxonomic hold-out benchmarks, ViraClass outperforms sequence-based and genome-content baselines. The largest gains emerge at deep evolutionary distances, in benchmarks that withhold entire families, orders or classes from the reference, where sequence-based methods lose most of their signal. In challenging boundary cases such as the *Flaviviridae*, ViraClass's structure-based placements capture the taxonomic boundary tensions highlighted by recent phylogenetic studies. When applied to a large collection of previously unclassified RdRp sequences, ViraClass places high-confidence queries into existing phyla and organizes the remainder into compact structural groups. ViraClass therefore provides a scalable approach from large-scale virus discovery to hierarchical taxonomic interpretation, particularly at the deep evolutionary ranges that current sequence-based pipelines cannot reach.

## Main

The known diversity of RNA viruses has expanded dramatically over the past decade, driven by large-scale metatranscriptomic surveys of environmental, host-associated and clinical samples[1–6]. Although this revolution in virus discovery is providing a new perspective on RNA virus evolution, it has also outpaced formal taxonomic assignment. The Master Species List 40 released by the International Committee on Taxonomy of Viruses (ICTV) recognizes seven phyla and 23 classes within the kingdom *Orthornavirae*[7–9], yet only a small fraction of the viral sequences assembled from environmental metatranscriptomic data have been placed within this hierarchy. Key obstacles are that the RNA-dependent RNA polymerase (RdRp) gene used for virus classification is typically the only gene universally carried by RNA viruses, and the low levels of RdRp amino acid sequence similarity among highly divergent viruses make many of them effectively undetectable by homology-based methods [10–12]. Consequently, taxonomic placement above the family level is often challenging, and sometimes impossible, from sequence evidence alone.

Several computational frameworks have been developed to address this problem. Sequence-based and profile-based pipelines, including VIRify[13], VirTAXA[14], RdRpBin[15] and VITAP[16], use homology or hidden Markov profiles of marker genes for detection or taxonomic prediction. Deep learning approaches such as CHEER[17] and PhaGCN2[18] extend sequence-based placement to short contigs or reads. Frameworks that operate on a genomic scale, including geNomad[19], VPF-Class[20] and vConTACT[21,22], incorporate marker annotation, protein content or gene-sharing relationships to refine placement beyond single-gene evidence. These methods have been instrumental for virus discovery, annotation and taxonomic placement, but they remain constrained when sequence homology is low or genome-wide protein content evidence is sparse, as is common for fragmented assemblies and RdRp-only data sets.

Protein structure preserves evolutionary signal on timescales over which sequence similarity has already collapsed[23]. This property is particularly relevant for the RdRp, whose right-handed polymerase architecture and conserved palm subdomain are maintained across the RNA virosphere[12,24]. Advances in accurate structure prediction through AlphaFold 3[25] and in database-scale structural search through Foldseek[26] have made it feasible to exploit this conserved signal at the scale of thousands of viruses. Structural comparison alone, however, is not sufficient for hierarchical taxonomic placement, as a useful framework must also decide when to accept, stop or defer placement, as well as handling queries that fall outside the reference space.

Herein, we present ViraClass, a structure-guided framework for the hierarchical taxonomic placement of RNA viruses from phylum to genus. ViraClass compares a predicted RdRp structure from a query virus against a reference set, determines at each taxonomic rank whether the structural evidence is sufficient for placement, and uses gene-sharing information to refine uncertain placements when genome-scale protein content is available. Queries that cannot be confidently placed into existing taxa are grouped into rank-calibrated structural clusters, thereby separating unplaced sequences into coherent candidate lineages. We first establish that RdRp structure carries taxonomic signal at depths where sequence similarity has collapsed, and that the organization of this signal recapitulates the current ICTV *Orthornavirae* classification. We then evaluate ViraClass across benchmarks with increasing difficulty, from randomly held-out species to entire orders and classes withheld from the reference. For the family *Flaviviridae*, whose phylum-level placement remains under active debate, ViraClass produced directional, family-wide structural placements toward *Pisuviricota*, contributing structural evidence at this contested phylum boundary[4,24]. Finally, we apply ViraClass to the unclassified RdRps recovered by a recent large-scale survey of the hidden RNA virosphere[5] and show that the majority of high-confidence sequences can be placed into existing phyla, with the remainder forming compact, well-separated structural groups comparable to established reference phyla.

## Results

### Data preparation and quantitative characterization of RdRp structural signal

To evaluate whether RdRp structure carries taxonomic signal beyond primary sequence, we assembled a reference set of RdRp structures spanning ICTV-recognized diversity within the kingdom *Orthornavirae*. Starting from the ICTV Virus Metadata Resource release associated with Master Species List 40 (VMR40)[7], we retained virus records assigned to *Orthornavirae* and linked to at least one GenBank nucleotide entry. RdRp regions were extracted from the viral genomes with Palm_annot [1], and structures of the RdRp sequences were predicted with AlphaFold 3[25]. This procedure yielded 7,256 viruses with both an RdRp sequence and a predicted RdRp structure (**Fig. 1a**). The resulting RdRp sequences concentrated in the expected size range of palm-domain-containing polymerases, with a mean length of 378.5 amino acids, and the predicted structures were generally of high confidence, with a mean pLDDT of 90.6 (**Fig. 1b**). This data set therefore provides a consistent substrate for systematic comparison of sequence-based and structure-based taxonomic signal.

Pairwise RdRp similarity across this set was quantified using one sequence-based measure – amino-acid sequence identity – as well as three structure-based measures: TM-score[27], local Distance Difference Test (lDDT)[28] and structure-aligned sequence identity. Sequence and structure comparisons were computed with MMseqs2[29] and Foldseek[26], respectively. To examine how RdRp similarity changes across the RNA virus hierarchy, virus pairs were grouped by their lowest shared taxonomic rank, together with pairs from different phyla. **Fig. 1c** shows that all four similarity measures separated same-genus pairs from other pairs well, but diverged markedly at deeper ranks. Sequence identity was compressed toward low values, such that class-, phylum- and cross-phylum comparisons occupied strongly overlapping ranges. In contrast, the three structure-based measures retained a clearer rank-dependent decrease from genus to cross-phylum comparisons. This difference was recapitulated by AUROC analysis (**Fig. 1d**), in which pairs sharing a given rank were treated as positives and all remaining pairs as negatives. lDDT gave the highest AUROC from phylum to family, reaching 0.940, 0.980, 0.983 and 0.992 at the phylum, class, order and family levels, respectively, whereas structure-aligned sequence identity was marginally higher at the genus level, with an AUROC of 0.991 compared with 0.988 for lDDT. In contrast, the AUROC for sequence identity declined from 0.976 at the genus level to 0.784 at the phylum level. RdRp structure therefore retains taxonomic signal across a broader range of evolutionary divergence than sequence alone.

This taxonomic signal was also reflected in the global organization of RdRp structural space. After restricting the data set to high-confidence structures with mean pLDDT $> 75$ and RdRp length $> 300$ amino acids, 6,684 structures remained for average-linkage hierarchical clustering based on pairwise lDDT distance (**Fig. 1e**). The resulting distance tree was separated into broad partitions corresponding to the six represented *Orthornavirae* phyla, including *Lenarviricota*, *Negarnaviricota*, *Duplornaviricota*, *Pisuviricota*, *Kitrinoviricota*, and *Ambiviricota*. Many lower-rank taxa also formed coherent subtrees within these partitions. To quantify agreement with current ICTV taxonomy, each taxon was scored as monophyletic when all of its members formed a single contiguous clade in the tree (**Fig. 1f**). Monophyly was observed for 3 of 6 phyla, 16 of 21 classes, 24 of 31 orders, 79 of 97 families, and 427 of 602 genera in the filtered set. RdRp structural space thus recovered most currently recognized taxa as compact hierarchical groups.

[1] https://github.com/rcedgar/palm_annot

**ViraClass enables accurate hierarchical taxonomic placement**

Guided by the taxonomic signal retained in the RdRp structure, we developed ViraClass as a hierarchical framework for RNA virus placement from phylum to genus (**Fig. 2a,b**). The final ViraClass algorithm was selected through an agentic exploration process that autonomously generated and evaluated candidate designs, retaining those that improved classification accuracy, coverage and calibration (**Fig. 2a, Supplementary Note 1**). Stage 1 uses the RdRp alone. The query RdRp is identified from the genome sequence, its structure is predicted with AlphaFold 3[25], and the resulting model is compared with a reference set of RdRp structures by structural search[26]. Placement then proceeds rank by rank, descending hierarchically from phylum to genus. At each step, candidate references are restricted to viruses consistent with the accepted higher-rank prefix, and the query is assigned by weighted nearest-neighbour voting in the corresponding structural distance space. To avoid overconfident placement, ViraClass calibrates acceptance thresholds separately at each rank using the support for the leading taxon and its margin over the next-best alternative. Hierarchical descent stops when either confidence measure falls below its calibrated threshold, so that each query is placed only to the deepest rank supported by the structural evidence.

When genome-wide protein information is available, ViraClass applies a second stage to queries that remain unresolved after structure-guided placement. This stage builds a gene-sharing network from reference remote protein families, represents each virus by the set of protein families detected in its genome, and compares the query with reference viruses within the taxonomic prefix already supported by stage 1. References sharing more protein families with the query receive higher weight, allowing ViraClass to refine the placement to a deeper rank when genome-wide gene content supports a consistent taxon. Queries that remain outside the known reference space after both stages are not forced into existing taxa. Instead, they are organized by hierarchical structural clustering using cut heights calibrated against the current ICTV ranks. ViraClass therefore supports three outcomes within a single workflow: placement in an existing lineage, additional resolution from genome-wide gene content, and rank-calibrated grouping of deeply divergent viruses that cannot yet be placed with confidence.

We next asked whether this design improves placement of previously unseen species while preserving reliability. We randomly withheld 30% of *Orthornavirae* species in the VMR40 dataset as a shared query set and evaluated ViraClass together with three representative methods, VITAP, geNomad and vConTACT3, across the genus, family, order, class and phylum levels. The overall accuracy was calculated across all query viruses at a given rank, such that unresolved or rejected queries reduced this metric. Coverage was defined as the fraction of queries receiving an accepted assignment at that rank. Precision was defined as the fraction of accepted assignments that were correct. ViraClass achieved the highest overall accuracy at the genus, family and order levels, where the benchmark stresses fine-grained discrimination among many closely related references (**Fig. 2c**). At the class and phylum levels, performance across methods converged. This convergence reflects the relatively lower difficulty of higher-rank placement under random species hold-out, in which close relatives of each query typically remain in the reference, so high-rank discrimination saturates for all methods. The deeper-placement advantage of ViraClass, its principal design goal, is therefore tested more directly in later clade hold-out analyses. ViraClass also maintained broad coverage across the hierarchy, with accepted placements for most queries from family to phylum and a substantial fraction at genus (**Fig. 2d**). Precision among accepted placements remained high across all ranks (**Fig. 2e**).

The baseline methods showed different trade-offs. VITAP also maintained high precision, but its lower coverage reduced overall accuracy, especially at the lower ranks. geNomad and vConTACT3 showed broader declines in both coverage and overall accuracy on this RNA virus benchmark. These patterns are consistent with the design of ViraClass. Structure-guided placement provides a strong signal when only RdRp is available, calibrated acceptance limits over-assignment, and the optional gene-sharing stage adds resolution when fuller genome information is present. Together, these results show that ViraClass can place previously unseen RNA virus species across the current taxonomic hierarchy while maintaining both high precision and broad practical coverage.

## ViraClass generalizes to newly added viruses

A practical placement framework must remain reliable as the reference taxonomy itself grows. To evaluate this property, we used the previous ICTV Virus Metadata Resource release, VMR39, as a fixed reference set and species first introduced in VMR40 as prospective queries. The reference set contained 6,955 viruses with available RdRp structures, and the query set contained 293 newly added species after excluding 14 entries without phylum annotation. The queries spanned three degrees of taxonomic novelty, namely 104 species placed in genera already represented in VMR39, 166 placed in genera that were themselves new but assigned to existing families, and 23 placed in families that were new but assigned to existing higher taxa. Each rank was therefore evaluated only on the subset of queries for which the true taxon was already present in VMR39, yielding 104 evaluable queries at genus, 270 at family, and all 293 at the order, class and phylum levels.

ViraClass placed newly added viruses substantially more accurately than VITAP at every rank (**Table 1**). The largest gains were observed at the family and genus ranks. At the family level, ViraClass exceeded VITAP by 35.9%, and at the genus level by 49.0%. This improvement reflected substantially broader accepted placement by ViraClass. Coverage exceeded that of VITAP at every rank, with the largest absolute gains again at family (88.1% vs 50.7%) and genus (73.1% vs 20.2%). Among accepted placements, both methods retained high reliability, with ViraClass precision between 94.7% at genus and 99.6% at phylum, and VITAP reaching 100.0% across all ranks. However, this VITAP precision is computed over only the small subset of queries that VITAP committed to placing, fewer than one in three at genus and roughly half at family, whereas ViraClass committed on most queries with comparably high precision. Together, these results indicate that ViraClass can be deployed prospectively across an ICTV update cycle, and that the additional coverage was concentrated in the query categories for which placement is most difficult.

## Structure-guided placement remains accurate beyond close relatives

A more demanding test of any placement framework is whether it remains informative when the close relatives of a query are absent from the reference set. We therefore constructed three benchmarks with progressively greater difficulty by withholding entire families, orders or classes from the VMR40 reference and using the withheld viruses as queries. For each benchmark, evaluation was restricted to ranks above the withheld clade, so that the task measured higher-rank placement beyond the removed rank.

ViraClass placed divergent queries more accurately than VITAP across all three settings (**Fig. 3a**). The advantage was largest when the missing reference space was broad. When the entire orders were withheld, ViraClass retained phylum-level accuracy of 96%, compared with 12% for VITAP; when entire classes were withheld, phylum-level accuracy was 67%, compared with 27% for VITAP. In the leave-one-family-out setting, ViraClass also achieved higher accuracy at order, class and phylum levels, reaching 72%, 84% and 95%, respectively, compared with 30%, 20% and 30% for VITAP. These results indicate that RdRp structure preserves placement signal across evolutionary distances where close sequence-level evidence is no longer available.

Coverage showed an even larger separation between the two methods (**Fig. 3b**). In the leave-one-order-out benchmark, ViraClass accepted placements for 91% of the queries at class level and 96% at phylum level, whereas VITAP accepted only 13% at both ranks. In the leave-one-class-out benchmark, phylum-level coverage was 74% for ViraClass and 34% for VITAP. A similar pattern was observed when families were withheld. These results suggest that removing close relatives reduces the evidence available for confident placement by sequence-based approaches.

### A *Flaviviridae* boundary case reveals taxonomic tension

The leave-one-class-out benchmark also exposed a boundary case of broader interest. In the current ICTV framework, the family *Flaviviridae* is placed within the *Kitrinoviricota*, class *Flasuviricetes*[8,9,30]. This placement has remained sensitive in large-scale RNA virus phylogenies[3,4]. In the analysis of Neri *et al.*, *Flasuviricetes* was recovered within the *Pisuviricota* and its phylum-level position was discussed as unresolved[4]. A recent structure-guided study of the family further documented the phylogenetic coherence of the *Flaviviridae* alongside a complex history of divergence and gene acquisition[24].

The same tension was reproduced in our experiment. When *Flasuviricetes* was withheld from the reference set, 91 of 144 *Flaviviridae* queries were placed in *Pisuviricota*, three were placed in *Negarnaviricota*, and 50 remained unresolved (**Fig. 3c**). The three *Negarnaviricota* placements corresponded to low-quality structures, while the dominant shift toward *Pisuviricota* was observed across *Pestivirus*, *Pegivirus*, *Hepacivirus* and *Orthoflavivirus*, suggesting that the signal was family-wide and not driven by any individual genus.

The structural distances were consistent with this placement pattern. The closest retained reference classes to *Flaviviridae* queries formed a mixed *Kitrinoviricota-Pisuviricota* neighbourhood, with representative nearest classes such as *Tolucaviricetes* and *Duplopiviricetes* showing mean lDDT distances of approximately 0.47-0.50 (**Fig. 3d**). An average-linkage tree built from the same distances placed the *Flaviviridae* cluster within the same compressed structural region near the phylum boundary (**Fig. 3e**).

These observations support a boundary-associated interpretation of the *Flaviviridae*. The placements were directional, family-wide and concentrated in a structurally close region of the RNA virus taxonomy whose phylum-level boundary has independently been reported as sensitive to taxon sampling and reference composition[4,24]. Thus, the *Flaviviridae* case suggests that structure-guided placement can expose taxonomic boundary tension under controlled reference perturbation, especially when the query virus occupies a region where current higher-rank assignments remain difficult to resolve.

### Real-world placement of previously unclassified LucaProt RdRps

To assess the ability of ViraClass to interpret RNA virus dark matter at scale, we applied it to the LucaProt collection of RdRps labelled "Unclassified by ICTV" or "New discoveries from this study"[5]. Among 161,981 LucaProt RdRps, 131,024 were already linked to known viruses, whereas 30,957 representative RdRps belonged to the unclassified and new-discovery subsets selected for downstream analysis (**Fig. 4a**). Structural confidence varied substantially across this set, and phylum-level placement into existing taxa increased with mean pLDDT, reaching its highest values in the high-confidence range (**Fig. 4b**). We therefore focused subsequent analyses on RdRps with mean pLDDT greater than 70, which enriched for reliable structural models while preserving a large set of previously unclassified sequences.

Within this high-confidence subset, ViraClass resolved many queries at the phylum level, whereas direct placement into existing classes remained limited. At the phylum level, 51% of the queries were placed into existing taxa and 49% were assigned to novel structural clusters. At the class level, 7% were placed into existing taxa, 92% were assigned to novel structural clusters, and 1% remained unresolved as singleton clusters (**Fig. 4c**). The contrast follows from the nature of the queries and the current reference landscape. The LucaProt subsets analyzed here are those for which the original survey did not obtain a confident assignment[5]. Such sequences are expected to be distant from labelled reference species, allowing broad phylum-level structural neighbourhoods to remain detectable while leaving class-level labels sparsely represented. Among accepted phylum-level placements, queries mapped predominantly to *Kitrinoviricota*, *Lenarviricota* and *Pisuviricota*, with a smaller share assigned to *Negarnaviricota* (**Fig. 4d**).

Queries that did not receive a confident placement into existing taxa were processed by the rank-calibrated clustering stage of ViraClass and organized into novel structural groups. The size distribution of these phylum-level novel clusters was broad, with cluster membership ranging from a few sequences to more than a hundred, while class-level clusters were on average smaller and more fragmented (**Fig. 4e**). To assess whether these phylum-level clusters were comparable in structure to established phyla, we examined two geometric properties of each cluster: the mean pairwise structural distance among its members and the structural distance from the cluster to its nearest known phylum. The first property quantifies internal coherence, with smaller values corresponding to more compact groups; the second quantifies separation, with larger values corresponding to clusters that are more clearly delimited from existing references. Established phyla occupied a characteristic region of this two-dimensional space, in which compact internal organization coexisted with substantial separation from neighbouring phyla. The novel phylum-level clusters identified by ViraClass occupied the same region (**Fig. 4f**). This analysis demonstrates that ViraClass extends hierarchical placement to a substantial portion of the RNA virus dark matter recovered at scale, while marking the remainder with structural groupings that can be examined by future taxonomic refinement.

## Discussion

The deep taxonomic placement of RNA viruses is limited by the rapid loss of RdRp sequence similarity, which leaves many newly recovered viruses difficult to classify. Here, we show that RdRp structure retains taxonomic signal across the RNA virosphere and that this signal can support hierarchical placement from phylum to genus. As such, we identify protein structure as a useful source of higher-rank taxonomic information in cases where sequence comparison is weak or genome context is incomplete. In the context of continuing ICTV expansion and large-scale metatranscriptomic discovery, ViraClass provides a practical approach for connecting newly recovered RNA viruses to hierarchical taxonomic interpretation[5,6,8,9].

ViraClass complements existing virus annotation and placement tools by focusing on cases in which the most reliable evidence is a divergent RdRp. Sequence- and profile-based frameworks rely on detectable sequence homology to reference markers, which becomes weak at deep ranks[12,16]. Genome-content frameworks use broader protein content and gene-sharing relationships[19,22], but these signals are often unavailable for partial RdRp-bearing contigs from metatranscriptomic assemblies. RdRp structure can remain conserved beyond the range of sequence similarity, as shown in the analyses of individual RNA virus families[12,24]. Recent advances in structure prediction and database-scale structural search now make it possible to use this signal at larger scale[25,26]. ViraClass converts RdRp structural similarity into a calibrated hierarchical workflow that can place a query into an existing taxon, stop placement when deeper-rank evidence is insufficient, or organize unplaced queries into rank-calibrated structural groups. The increasing advantage of ViraClass in deeper clade-holdout benchmarks is consistent with the slower loss of structural similarity compared with sequence similarity across large evolutionary distances.

Several limitations define the current scope of the framework. First, ViraClass depends on the accurate identification of RdRp sequences and the quality of the predicted RdRp structures. This dependence was evident in LucaProt, where placement into existing phyla increased with pLDDT and low-confidence structures contributed to unstable placements in difficult cases such as *Flaviviridae*. Because high-quality structure prediction requires a correctly identified and sufficiently complete RdRp sequence, improved RdRp detection from viral genomes and assemblies, together with better structure modelling for highly divergent RdRps, should increase coverage and placement stability in these regions. Second, ViraClass remains dependent on the coverage and balance of the reference database. Structure-guided placement can only recover relationships represented by current reference structures. Reference imbalance can also affect nearest-neighbour voting because large taxa contribute more candidate neighbours than small taxa. Expanding the reference database with more evenly sampled RNA virus lineages and more complete genomes should improve direct placement at finer ranks. Third, benchmark evaluations based on the ICTV labels inherit the history of the taxonomy itself, which has incorporated substantial sequence-based evidence in defining higher-rank boundaries. Independent evaluation against future taxonomic revisions, phylogenetic analyses and ecological coherence will therefore remain important.

The LucaProt analysis illustrates how structure-guided placement can help interpret RNA virus dark matter discovered at scale. ViraClass placed many previously unclassified RdRps into existing phyla and organized much of the remaining diversity into compact structural groups with internal similarity and nearest-reference separation comparable to established phyla. These results suggest that a large fraction of unclassified LucaProt RdRps contains coherent higher-rank structure that is poorly represented by the current taxonomy framework. More broadly, the LucaProt results indicate that phylum-level structural neighbourhoods are partly covered by current references, whereas class-level labels remain sparse for many deeply divergent RdRps. The *Flaviviridae* case further shows that structure-guided placement can highlight specific taxonomic groups where current higher-rank boundaries are sensitive to reference composition and taxon sampling. ViraClass therefore provides both a placement framework for new RNA viruses and a structural view of where the current taxonomy is well sampled, under sampled or locally uncertain.

## Methods

### Data sources and reference taxonomy

The reference taxonomy was derived from the ICTV Virus Metadata Resource (VMR), the curated registry that accompanies each ICTV Master Species List (MSL). Two consecutive releases were used. VMR_MSL40 v2 (hereafter VMR40) served as the primary reference for the closed-set benchmarks, the threshold-calibration analyses and the LucaProt application. VMR_MSL39 v4 (hereafter VMR39) was used as a temporal reference to evaluate prospective generalization across an ICTV update cycle. From each release we retained virus records assigned to the kingdom *Orthornavirae*, linked to at least one GenBank nucleotide accession, and associated with a curated canonical RdRp representative sequence. After this filtering, the VMR40 reference contained 7,256 viruses with both an RdRp sequence and a predicted RdRp structure, and the VMR39 reference contained 6,955 such viruses.

**Protein product extraction**

Virus-level protein sequences were obtained from GenBank records. All annotated coding sequence (CDS) features were extracted; when a CDS translation was absent, the nucleotide feature was translated using the recorded transl_table. For accessions without any annotated CDS, open reading frames were predicted with orfipy v0.0.4[31] using min_nt = 90, strand = both, table = 1, case-insensitive parsing and –single-mode. Mature peptide annotations (mat_peptide or mat_peptide_nt) were prioritized over their parent CDS products.

To better represent the modular architecture of RNA virus polyproteins, candidate proteins of length ≥ 1,000 amino acids were further analysed with InterProScan 5[32]. InterPro matches were mapped to six hallmark categories: protease, helicase, RdRp, methyltransferase, capsid or nucleocapsid, and envelope glycoprotein. A protein was split only when it was at least 1,000 amino acids long and contained at least two non-overlapping hallmark domains. Mature peptides were emitted as independent protein units, unsplit CDS products were retained at full length, and split proteins were emitted as one unit per hallmark domain. The resulting per-virus protein set was used as the input for the gene-sharing analysis described below.

**RdRp identification, structure prediction and structural comparison**

Candidate RdRp regions were identified from full-length viral nucleotide sequences with Palm_annot (https://github.com/rcedgar/palm_annot) under default parameters. The Palm-domain–derived RdRp amino acid sequences were used as input for structure prediction with AlphaFold 3[25], and the resulting predicted RdRp structures were retained as the structural representation of each virus. Pairwise structural similarities were computed with Foldseek[26] (foldseek easy-search –alignment-type 2 -e inf –exhaustive-search 1), retaining TM-score, lDDT and structure-aligned sequence identity. Similarity matrices were symmetrized, normalized and converted to distances as $d = 1 - s$. LDDT-derived distances were used from phylum to family, and structure-aligned sequence identity-derived distances were used at genus, where local sequence-level resolution was more informative.

**Overview of ViraClass**

ViraClass is a calibrated two-stage hierarchical classifier that places a query RNA virus across the five ICTV ranks from phylum to genus. Stage 1 places each query using RdRp structure alone, by applying distance-weighted $k$-nearest-neighbour voting to pairwise structural distances between the query and reference RdRps. Stage 2 is invoked only for ranks that Stage 1 leaves unresolved and only when a genome-level protein catalog is available; it refines or rescues placements using a gene-sharing similarity derived from remote protein families. Acceptance at each rank is governed by separately calibrated thresholds, so descent stops at the deepest rank supported by the evidence. Queries that remain unresolved after both stages can be organized into rank-calibrated structural clusters by hierarchical clustering of their lDDT distances, providing provisional groupings outside the existing reference taxonomy. The two stages and the clustering layer are described in turn below.

**Stage 1: structure-guided hierarchical classification**

Stage 1 traverses the ICTV ranks $r \in \{\mathrm{Phylum}, \mathrm{Class}, \mathrm{Order}, \mathrm{Family}, \mathrm{Genus}\}$ in top-down order. Before voting, a global distance gate was applied: a query whose minimum lDDT distance to all references exceeded 0.45 was rejected at every rank as too structurally distant from the reference set to support placement. At each rank $r$, candidate references were restricted to viruses consistent with all previously accepted parent-rank predictions (prefix constraint). Placement was then performed by weighted $k$-nearest neighbors on the structural distance, using $k = \{120,80,30,3,3\}$ from phylum to genus. Neighbor $i$ voted for its rank-$r$ taxon with weight

$$w_i = \frac{1}{d_i + \varepsilon},$$

where $d_i$ is the query-to-neighbor distance and $\varepsilon = 10^{-8}$ is a small constant introduced to avoid division by zero. The predicted taxon was the one with the largest summed vote.

Two decision statistics were recorded at each rank: vote confidence,

$$c = \frac{s_1}{\sum_j s_j},$$

and vote margin,

$$m = s_1 - s_2,$$

where $s_1$ and $s_2$ are the largest and second-largest taxon vote totals. A rank was accepted only if both $c \geq T_r$ and $m \geq M_r$. Rejection at a given rank terminated supervised descent below the accepted prefix. Phylum-, class-, order- and family-level placement used the lDDT distance matrix. Genus-level placement used structure-aligned sequence identity-derived distance matrix.

**Stage 2: gene-sharing network refinement**

Stage 2 was designed to refine or rescue assignments that remained unresolved after structure-guided placement in stage 1. It used remote protein families derived from the reference genomes to quantify gene-sharing relationships among viruses. Reference proteins were first clustered with MMseqs2 easy-cluster[29] at 30% amino-acid identity (--min-seq-id 0.30 -c 0.8 --cov-mode 0 --cluster-mode 2) to define protein clusters (PC30). PC30s present in at least two reference viruses were treated as core clusters; all others were treated as rare. For each core PC30, protein sequences were aligned with MAFFT v7[33] and converted to profile hidden Markov models with HMMER3[34]. Reciprocal HMM hits passing the specified E-value and coverage filters were converted to a weighted similarity graph, and remote protein families were obtained by Markov clustering with inflation 2.0[35]. Rare PC30s were then attached to the best matching core-derived remote family if they passed stricter HMMER acceptance criteria. A final frozen HMM database was built over the resulting remote protein families.

Query proteins were mapped to this reference HMM database using the same search framework. Each virus was then represented as a binary presence–absence vector over the remote protein-family vocabulary, with one dimension per remote family and a value of 1 indicating that at least one protein from the virus matched that family. For a query virus $q$ and reference virus $r$, stage-2 similarity was defined as a square-root Jaccard score,

$$\mathrm{sim}(q,r) = \frac{|R_q \cap R_r|}{\sqrt{|R_q|\,|R_r|}},$$

where $R_q$ and $R_r$ are the sets of remote protein families detected in the two viruses. Distance was defined as $d = 1 - \mathrm{sim}(q,r)$.

Stage 2 started below the deepest rank already accepted by stage 1 and applied the same prefix-constrained weighted $k$-nearest-neighbor rule, with $k = \{30,15,3,3,3\}$ from phylum to genus. To reduce unnecessary comparisons, candidate references were intersected with the top 25 structural neighbors from stage 1 within the currently allowed prefix.

**Threshold calibration**

Acceptance thresholds $T_c(r)$ and $T_m(r)$ were calibrated separately for each taxonomic rank and separately for Stage 1 and Stage 2. Calibration was performed on the VMR40 reference set using held-out taxa at six levels: species, genus, family, order, class and phylum. These hold-outs created two types of test cases for each rank $r$. In positive cases, the correct taxon at rank $r$ was still present in the reference set, so a correct placement was possible. In negative cases, the correct taxon at rank $r$ had been removed from the reference set, so any accepted placement at that rank would represent a false acceptance. Lower-rank hold-outs were sampled by a taxon-level split, whereas order-, class- and phylum-level hold-outs were evaluated by exhaustive leave-one-clade-out enumeration. Held-out taxa were included only when at least one sibling taxon remained under the same parent. To prevent large strata from dominating calibration, the six hold-out strata were given equal total weight.

For each held-out query, we computed two decision statistics: vote confidence $c$, defined as the fraction of total vote weight assigned to the leading taxon, and vote margin $m$, defined as the difference in vote weight between the leading and second-ranked taxa. Candidate threshold pairs $(T_c, T_m)$ were evaluated on a $60 \times 60$ grid over the observed values of $c$ and $m$. A threshold pair was retained if accepted positive cases reached at least 0.80 precision. For each rank, we additionally required that the weighted false-acceptance rate on negative cases not exceed a ceiling value. The default ceiling was set to 0.20. At the genus level, we relaxed this ceiling to 0.50. Among all retained threshold pairs, we selected the one that accepted the largest number of positive cases. If multiple pairs gave the same positive coverage, we chose the pair with the lower false-acceptance rate, followed by the more stringent thresholds.

**Rank-calibrated structural clustering**

To group queries that did not receive a confident placement into existing taxa, the lDDT distance matrix was clustered hierarchically with average linkage, and the dendrogram was cut at a rank-specific height calibrated against the current ICTV taxonomy. Cut heights were calibrated conditionally on the parent taxon: phylum-level cuts were calibrated globally, class-level cuts within each known phylum, order-level cuts within each known class, family-level cuts within each known order and genus-level cuts within each known family. For each rank, the cut height was selected to maximize the adjusted Rand index (ARI), which measures how closely structural clusters match the corresponding ICTV taxa, with 1 indicating perfect agreement and 0 indicating chance-level agreement. For unresolved queries, sequences sharing the same accepted prefix and the same first-unresolved rank were clustered together using the corresponding rank-specific cut height (**Supplementary Note 2**). Clusters with at least two members were reported as rank-calibrated novel taxa, whereas smaller groups were retained as singletons.

**Agent-guided exploration**

As shown in **Fig. 2a**, the final ViraClass workflow was selected through a benchmark-guided search over candidate classification strategies, automated by a large language model (LLM)-based agent system inspired by InternAgent[36,37]. We used this approach because hierarchical RNA virus placement involves several interacting choices: whether to rely on RdRp sequence, RdRp structure, genome-wide protein content, or a combination, how to score similarity between viruses, how nearest-neighbour information should be combined into a prediction, and when to accept or withhold a placement — and the best combination of these choices was not obvious in advance. In each round, a candidate placement workflow was generated, run end-to-end on the benchmark, and scored on accuracy, coverage, calibration and computational cost. Both successful and unsuccessful candidates were kept on record, so that the search could progressively converge on choices that worked reliably across taxonomic ranks rather than only at one level. The final ViraClass algorithm was then assembled from the best-performing candidates, with manual refinement to ensure that each component was biologically motivated and interpretable. Additional details and the full exploration trajectory are provided in **Supplementary Note 1**.

### Baseline methods

On the VMR40 leave-species-out benchmark, we compared ViraClass with VITAP[16], geNomad[19] and vConTACT3[22]. All baselines used the concatenated nucleotide FASTA of the held-out query genomes as input. geNomad was run directly on this query FASTA. VITAP was run on the same query FASTA against its default hybrid reference database (DB_hybrid_MSL37_RefSeq209_IMGVR). vConTACT3 was run with reference database v232, and queries were split into eukaryotic and prokaryotic subsets because the method uses domain-specific reference databases. Host groups were assigned from the ICTV Host source field; cases lacking a direct host annotation were resolved from species-level annotation. For the leave-family, leave-order and leave-class benchmarks, we compared ViraClass only with VITAP. In these settings, VITAP used the same reference-query split as ViraClass. The shared benchmark reference set was converted to the accession-level lineage format required by VITAP and indexed as a benchmark-matched custom reference database.

### LucaProt RdRp analysis

The LucaProt analysis used the subset of LucaProt RdRps corresponding to sequences labelled as "Unclassified by ICTV" or "New discoveries from this study"[5]. Because these entries were represented by RdRp sequences without genome-wide protein catalogs suitable for gene-sharing refinement, this analysis used the structure-guided placement and rank-calibrated structural clustering only. Downstream analyses focused on sequences with pLDDT $> 70$.

**Data availability**

All ICTV taxonomy and virus metadata used in this study are publicly available from the International Committee on Taxonomy of Viruses. The Virus Metadata Resource (VMR) is available at https://ictv.global/vmr, and the Master Species List (MSL) is available at https://ictv.global/msl.

**Code availability**

The ViraClass source code will be released in a public repository upon acceptance of the manuscript.

**Author contributions**

W.S. and S.S. conceived and supervised the project. S.X., W.H. and S.Y. conceptualized the study, preprocessed data, developed the framework, analysed the results, wrote the manuscript and prepared the figures. W.H., S.F., X.H. and B.Z. developed the agent and memory modules. W.B. and Q.F. performed RdRp structure prediction. E.C.H., W.S. and S.S. revised the manuscript. All authors read and approved the final manuscript.

**Acknowledgments**

This work was funded by the Prevention and Control of Emerging and Major Infectious Diseases-National Science and Technology Major Project (2025ZD01903700).

**Competing interests**

The authors declare no competing interests.

**Supplementary Information**

The supplementary information includes

- Supplementary Note 1: Detailed framework of the agentic exploration
- Supplementary Note 2: Calibration of rank-specific cut heights for hierarchical structural clustering
- Supplementary Figs. 1 to 3
- Appendix A: Prompts for agents in the discovery system
- Appendix B: Detailed agent exploration trajectory.

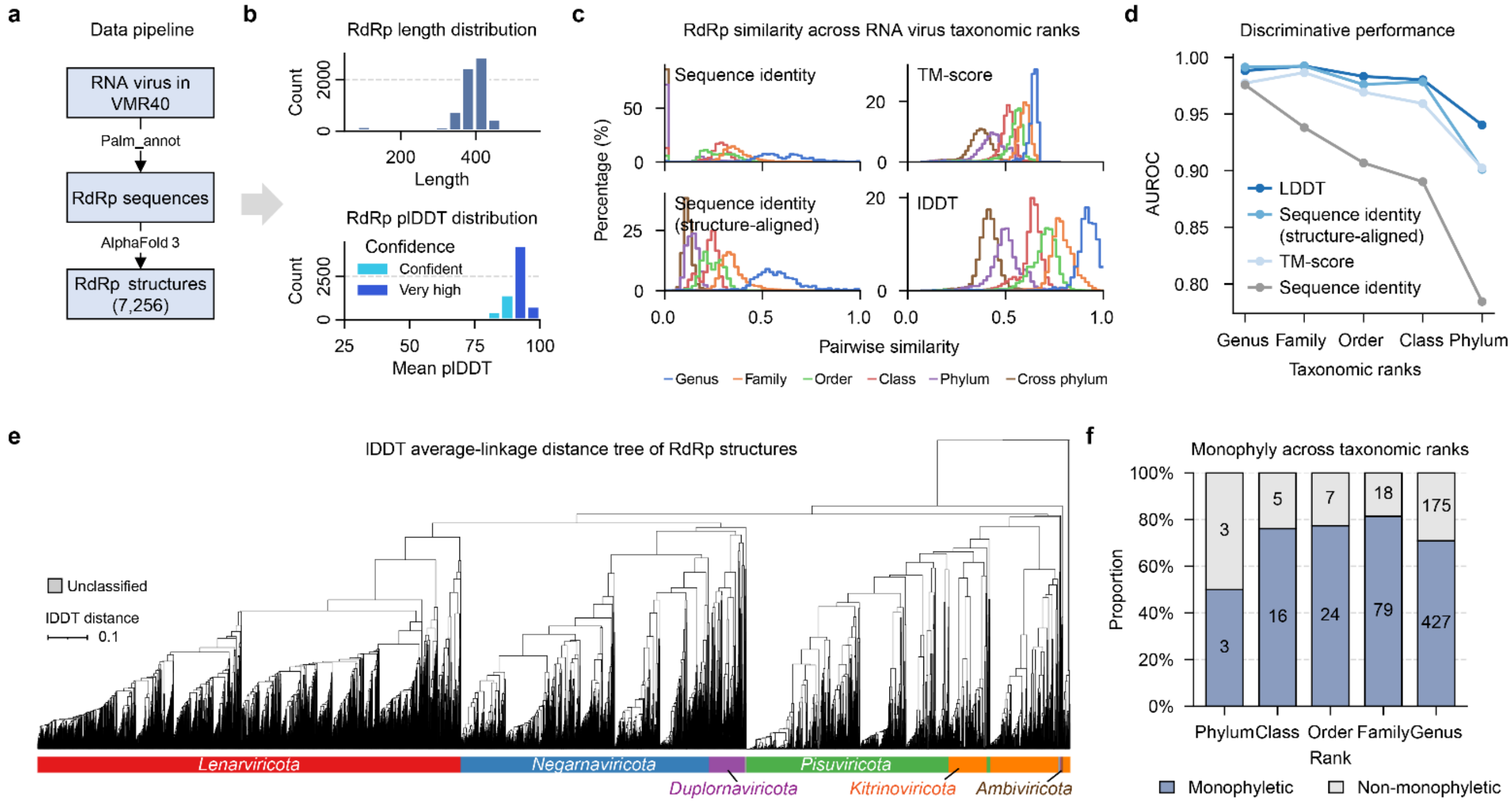

**Fig. 1: Quantitative characterizations of RdRp structural signal across the RNA virosphere. a,** Data preparation workflow. *Orthornavirae* entries from the ICTV Virus Metadata Resource release associated with Master Species List 40 (VMR40) and linked to GenBank nucleotide records were processed with Palm_annot to extract Palm-domain-derived RdRp sequences, followed by AlphaFold 3 structure prediction, yielding 7,256 virus records. **b,** RdRp length distribution and mean pLDDT of the predicted structures. **c,** Pairwise RdRp sequence and structure similarity across the lowest shared taxonomic ranks, from genus to cross-phylum comparisons. **d,** Discrimination of taxonomic relationships by similarity, quantified as AUROC for classification of whether a pair shares a given rank. **e,** Average-linkage clustering of 6,684 high-confidence RdRp structures using lDDT distance with mean pLDDT > 75 and length > 300 amino acids. Colours denote phyla. **f,** Proportion of taxa recovered as monophyletic in the lDDT tree across ranks. Numbers indicate counts of monophyletic and non-monophyletic taxa.

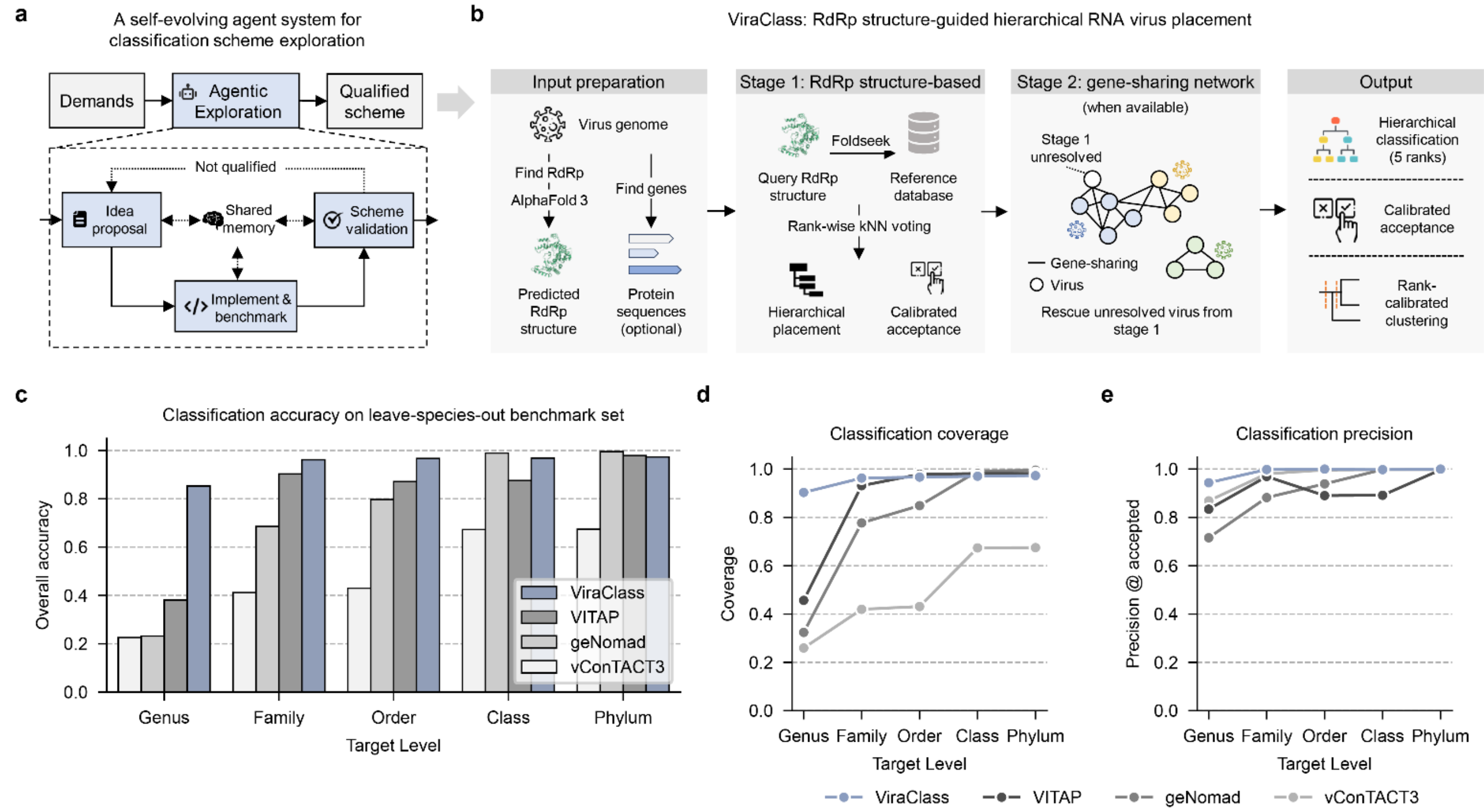

**Fig. 2: The ViraClass framework and performance on the leave-species-out benchmark. a,** The agentic exploration system used to identify the final ViraClass algorithm. **b,** Overview of ViraClass. Stage 1 performs RdRp structure-guided hierarchical placement with calibrated acceptance. Stage 2 uses a gene-sharing network, when genome-wide protein information is available, to refine or rescue queries unresolved after stage 1. Queries that remain outside the reference taxonomy are organized by rank-calibrated structural clustering. **c,** The overall taxonomic assignment accuracy on the leave-species-out benchmark, in which 30% of species, corresponding to 1,966 query virus records, in the VMR40 dataset were withheld as queries and evaluated from genus to phylum. **d,** Coverage, defined as the fraction of queries receiving an accepted assignment at each rank. **e,** Precision among accepted assignments at each rank.

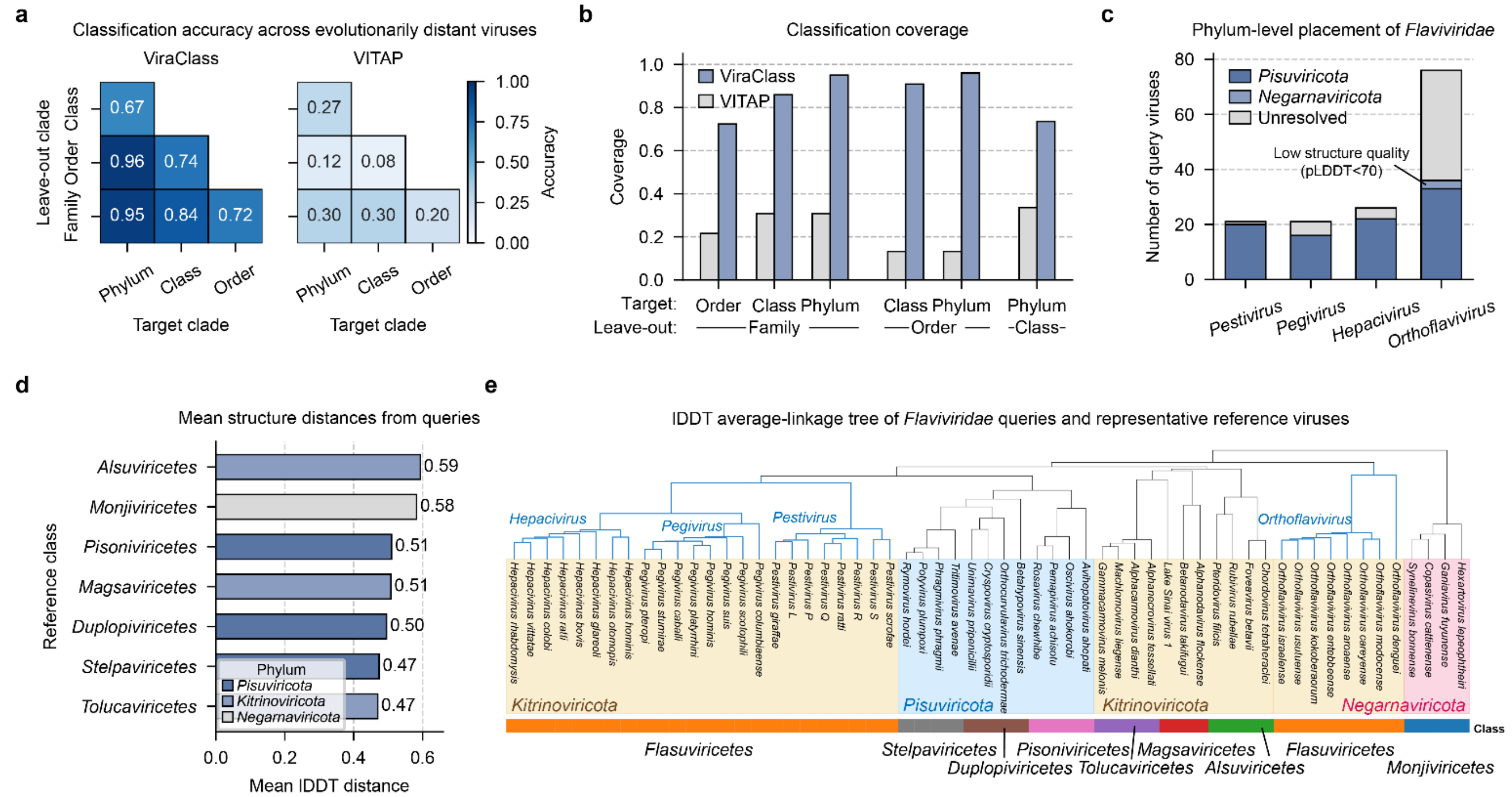

**Fig. 3: Structure-guided placement at deep evolutionary distances and the *Flaviviridae* boundary case. a,** The placement accuracy of ViraClass and VITAP under leave-one-family-out, leave-one-order-out and leave-one-class-out evaluation. The evaluation included 118 evaluable family-out folds (n = 6,461 query-fold records), 14 evaluable order-out folds (n = 6,111 query-fold records) and 19 evaluable class-out folds (n = 4,659 query-fold records). Predictions were assessed only at ranks above the withheld clade. Colour intensity indicates accuracy. **b,** The coverage across the same hold-out settings, defined as the fraction of evaluable queries for which a placement was accepted. **c,** Phylum-level placements of *Flaviviridae* queries when the parent class *Flasuviricete*s was withheld from the reference set, stratified by genus. Most accepted queries were placed in the *Pisuviricota*, and the small *Negarnaviricota* fraction corresponded to low-confidence predicted structures. **d,** Mean lDDT distances from *Flaviviridae* queries to reference classes. Bar colour indicates the phylum to which each reference class is assigned. **e,** Average-linkage tree of pairwise lDDT distances between representative *Flaviviridae* queries and neighbouring reference viruses. The *Flaviviridae* queries form a compact cluster whose nearest reference neighbourhood spans the boundary between *Kitrinoviricota* and *Pisuviricota*.

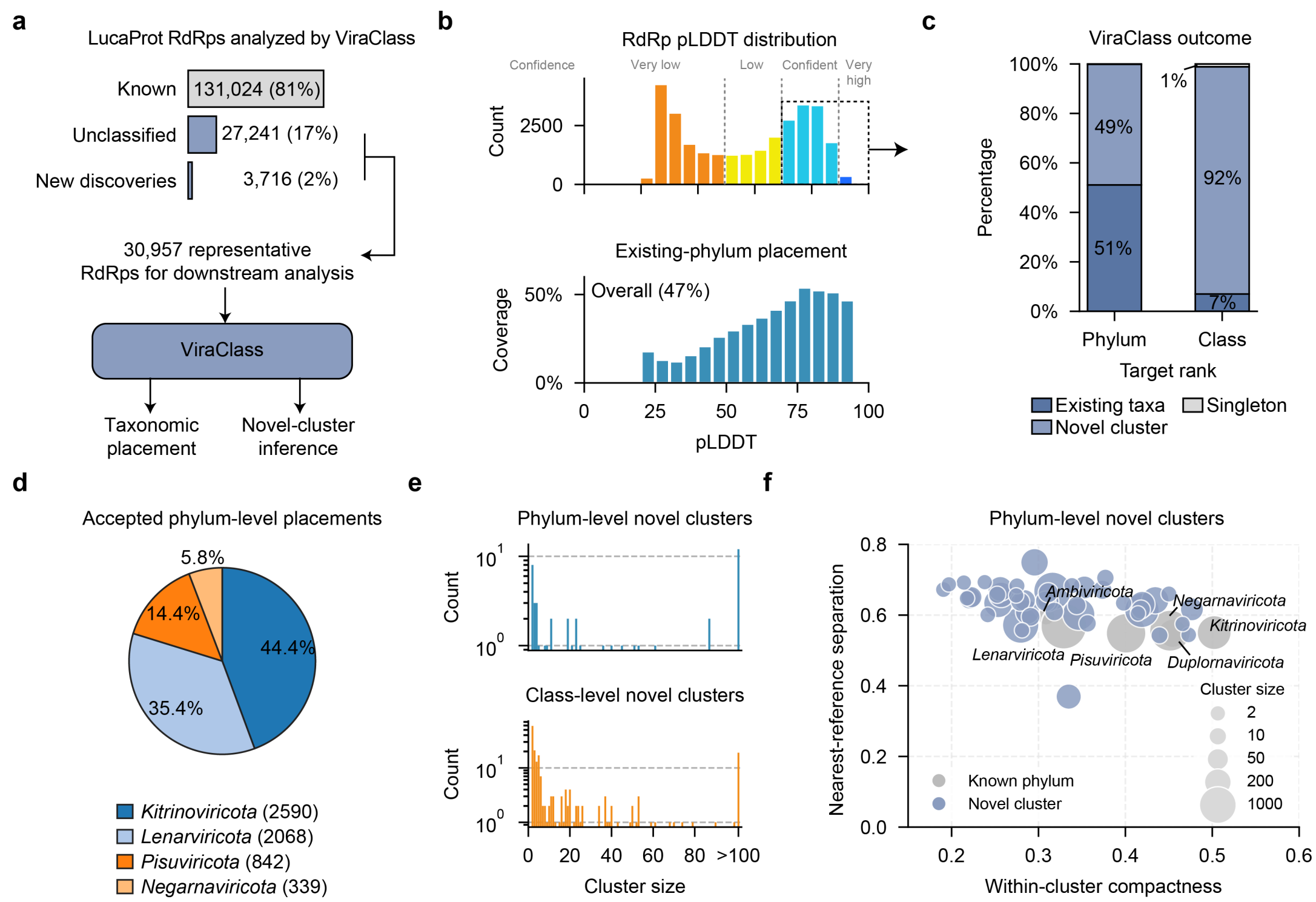

**Fig. 4: Real-world placement of previously unclassified LucaProt RdRps by ViraClass. a,** Composition of the LucaProt RdRp collection analyzed by ViraClass. Known viruses were separated from the Unclassified by ICTV and New discoveries from this study, yielding 30,957 representative RdRps for downstream analysis. **b,** Mean pLDDT distribution across the representative LucaProt RdRps and phylum-level placement into existing taxa as a function of structural confidence. Downstream analyses focused on sequences with mean pLDDT greater than 70. **c,** ViraClass outcomes for the high-confidence subset at phylum and class. Unresolved sequences correspond to singleton clusters. **d,** Distribution of accepted phylum-level placements among existing phyla. **e,** Size distributions of phylum-level and class-level novel structural clusters. **f,** Within-cluster compactness and nearest-reference separation for phylum-level novel clusters and known reference phyla. Point size indicates cluster size.

**Table 1: Prospective taxonomic placement from VMR39 to VMR40.** The VMR39 viruses were used as the reference set, and viruses newly introduced in VMR40 were used as queries. Values represent percentages. Parentheses indicate the number of evaluable viruses at each rank. Accuracy is the fraction of evaluable queries that were correctly placed, coverage is the fraction of evaluable queries for which a placement was accepted. Precision@accepted is the fraction of accepted placements that were correct, and is therefore evaluated on each method's accepted subset rather than on the full query set. Because VITAP's coverage is markedly lower than ViraClass's at every rank (for example, 50.7% versus 88.1% at family and 20.2% versus 73.1% at genus), VITAP's Precision@accepted values reflect performance on a much smaller fraction of queries than the corresponding ViraClass values. $\Delta$ denotes the absolute difference between ViraClass and VITAP.

| | **Accuracy** | | | **Coverage** | | | **Precision@accepted** | | |
|---|---|---|---|---|---|---|---|---|---|
| **Rank** | ViraClass | VITAP | Δ | ViraClass | VITAP | Δ | ViraClass | VITAP | Δ |
| Phylum ($n = 293$) | **91.8** | 73.7 | +18.1 | **92.2** | 73.7 | +18.4 | 99.6 | **100.0** | -0.4 |
| Class ($n = 293$) | **88.1** | 73.7 | +14.4 | **90.8** | 73.7 | +17.1 | 97.0 | **100.0** | -3.0 |
| Order ($n = 293$) | **85.7** | 68.6 | +17.1 | **89.1** | 68.6 | +20.5 | 96.2 | **100.0** | -3.8 |
| Family ($n = 270$) | **86.7** | 50.7 | +36.0 | **88.1** | 50.7 | +37.4 | 98.3 | **100.0** | -1.7 |
| Genus ($n = 104$) | **69.2** | 20.2 | +49.0 | **73.1** | 20.2 | +52.9 | 94.7 | **100.0** | -5.3 |